\documentclass[fleqn,12pt,twoside]{article}
\usepackage{espcrc1}

\usepackage{graphicx}
\usepackage[figuresright]{rotating}

\newcommand{\AmS}{{\protect\the\textfont2
  A\kern-.1667em\lower.5ex\hbox{M}\kern-.125emS}}
\title{Making carbon in stars}
\author{Sam M. Austin
\address{National Superconducting Cyclotron Laboratory and  \\
Joint Institute for Nuclear Astrophysics (JINA) \\
NSCL, Michigan State University, East Lansing MI 48824, USA\\
E-mail: austin@nscl.msu.edu}
\thanks{This work was supported in part by the US National Science
Foundation grants PHY01-10253 and PHY02-16783, the latter  funding
the Joint Institute for Nuclear Astrophysics (JINA), an NSF
Physics Frontier Center.}}

\begin{document}
\bibliographystyle{unsrt}
\maketitle

\begin{abstract}
The triple alpha ($3\alpha$) process plays an important role in
the production of $^{12}$C in stars. Its rate is known with an
accuracy of about 12\%. We examine the corresponding uncertainties
introduced in the description of pre-supernova stars, of
nucleosynthesis in a core-collapse SN explosion, and of the
production of $^{12}$C during the third dredge-up in asymptotic
giant branch (AGB) stars.  For the AGB case we consider also the effects
of uncertainties in the $^{14}$N($p, \gamma)^{15}$O rate.  We
conclude that the present accuracy of the $3\alpha$ rate is
inadequate and describe new experiments that will lead to a more
accurate value.
\end{abstract}

\section{INTRODUCTION}
Although the $3\alpha$ reaction plays the central role in the
production of $^{12}$C in stars, little attention has been paid to
the effect of uncertainties in its rate.  The emphasis has been on
the following reaction $^{12}$C($\alpha,\gamma$)$^{16}$O. During
core helium burning, these two reactions are in competition, and
the ratio R of their rates determines the relative amounts of
$^{12}$C and $^{16}$O produced. The C/O ratio in turn affects the
later evolution of the star.  In the case of the pre-supernova
evolution of a massive star, for example, it is found that the
size of the Fe core at the onset of core collapse depends on R, as
does the composition of the material later ejected into the
interstellar medium. In the context of a parametrized  explosion
model, the composition of the nucleosynthesized material
constrains R to within about 10\%. Given the required $\pm 10\%$
precision in R the present $\pm 12\%$ precision of the $3\alpha $
rate is inadequate.

The $3\alpha$ process also plays an important role in asymptotic
giant branch (AGB) stars, stars burning hydrogen and helium in
shells around a degenerate core containing mainly carbon and
oxygen.  We \cite{herwig04} have investigated $^{12}$C production
is these stars and have found that uncertainties in the
$^{12}$C($\alpha,\gamma$)$^{16}$O rate have little effect, but
that uncertainties in the $3\alpha$ and $^{14}$N($p,
\gamma)^{15}$O rates are important.

In the following sections we will discuss these phenomena in more
detail and will then discuss measurements that promise to improve
the accuracy of the $3\alpha$ reaction rate to about $\pm 6\%$.

\section{CORE-COLLAPSE SUPERNOVAE}
The dependence of the synthesis of A=16-40 nuclei on R is well
documented. Calculations were performed \cite{woosley03} for a $25
M_{sun}$ star using a range of values of the
$^{12}$C($\alpha,\gamma$)$^{16}$O reaction rate, $r_{\alpha,12}$,
and assuming a standard value of the $3\alpha$ rate. For values of
$r_{\alpha,12}$  of $1.2 (\pm 10\%)$ times the rate suggested by
Buchmann \cite{buchmann96}, the production factors for these
nuclei are almost independent of A, allowing a simultaneous
reproduction of their abundances.  Less well known is the
dependence of the size of the pre-collapse iron core on R.  Fig.~1
shows this dependence and the relevant range of R as determined
from the nucleosynthesis constraints  discussed above. The core
mass varies significantly as a function of R, by about
$0.2M_{sun}$ in the relevant range. This uncertainty is likely to
be important for the behavior of the SN explosion; it takes
$3\times10^{51}$ ergs to dissociate $0.2M_{sun}$ into nucleons,
similar to (or greater than?) the energy released in a supernova
explosion.

\begin{figure}[htb]
\begin{center}
\includegraphics[height=3.5in]{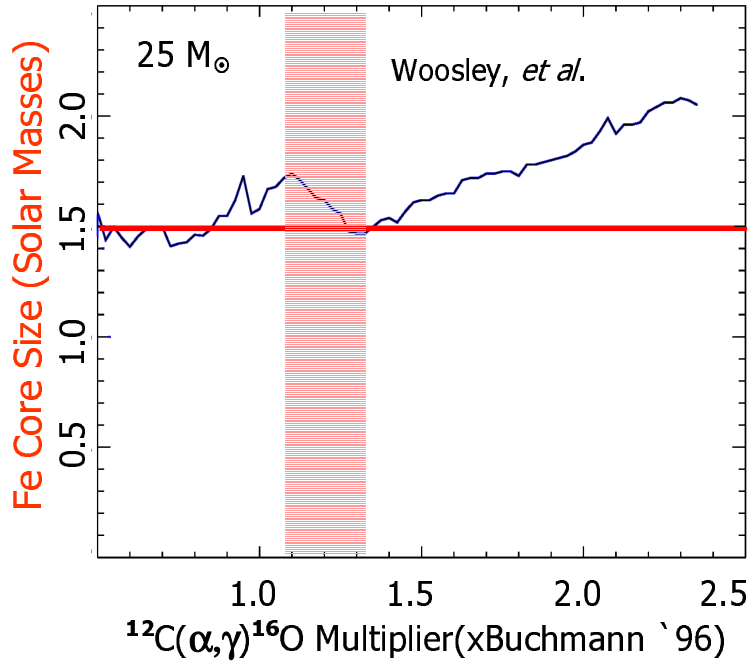}
\end{center}
\vspace{-0.4in} \caption{The Fe core size obtained
\cite{woosley03} for a $25M_{sun}$ pre-supernova star for various
values of the $^{12}$C($\alpha,\gamma$)$^{16}$O reaction rate(and
hence of R).  The unit rate is that recommended by Buchmann
\cite{buchmann96}. The shaded bar corresponds to the range of R
that gives a nearly constant production factor for $A= 16-40$
nuclei.}
\end{figure}

This is an important issue.  Even if R were determined with 10\%
accuracy, and that will not be easy, there would be  an intrinsic
uncertainty of $0.2M_{sun}$ in the Fe core mass for this star.  At
least two issues need to be examined.  First, whether the value of
R obtained from nuclear experiments is consistent with that
obtained from SN models of nucleosynthesis.  That is,  do the SN
models reflect what really happens in stars?  And second, whether
rapid core mass changes with R occur for a range of stellar
masses.

\section{AGB STARS}
Following completion of core hydrogen and helium burning in
intermediate mass stars, hydrogen and helium are burned in shells
surrounding a degenerate carbon-oxygen core.  Eventually, burning
in the helium shell becomes unstable and thermal flashes induce
convective behavior and production of $^{12}$C in the inter-shell
region. Later, the surface convection zone moves into the
inter-shell region; carbon-rich material enters the convective
envelope of the star and is carried to the stellar surface in a
process known as the third dredge-up. The surface of the star
becomes carbon rich and much of this material is blown into the
interstellar medium. Such behavior is observed for light stars,
2-3 $M_{sun}$ in the Magellanic clouds. However, it has not, so
far, been possible to reproduce this behavior theoretically. The
flash behavior found in most calculations is too weak to produce
enough $^{12}$C, although certain approximations involved in these
calculations prevent firm conclusions. We discuss here whether
uncertainties in the reaction rates involved might also have
important effects.
\begin{figure}[h]
\begin{center}
\includegraphics[height=3.0in]{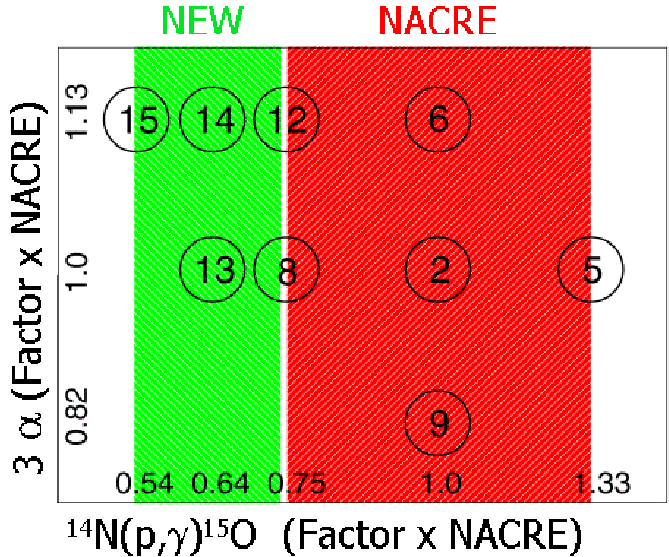}
\end{center}
\vspace{-0.4in} \caption{Values of the reaction rates at which
calculations have been done, compared to the NACRE values.  On the
right the circles are at the NACRE values $\pm$ their errors; on
the left are the new values of the
$^{14}$N($p, \gamma)^{15}$O rates. From F. Herwig.}
\end{figure}
It is known that stronger flashes lead to more efficient dredge-up
and that weaker hydrogen burning leads to stronger flashes, but
these phenomena have not been studied in a systematic fashion and
related to changes in the underlying nuclear reaction rates. We
describe here calculations made to study  the sensitivity of the
AGB-thermal flash process to the rates for the $3\alpha$,
$^{12}$C($\alpha,\gamma$)$^{16}$O, and $^{14}N(p, \gamma)^{15}$O
reactions. Additional details can be found in Ref. \cite{herwig04}.

Initially we used the NACRE \cite{angulo99} reaction rates and
their uncertainties, as shown in the right-hand area of Fig.~2. We
found that carbon production was insensitive to the
$^{12}$C($\alpha,\gamma$)$^{16}$O rate. However, changes in the
$3\alpha$ and  $^{14}$N($p, \gamma)^{15}$O  rates produced large
effects.  To estimate these effects quantitatively, one rate at a
time was changed to its upper or lower error limit.  The
calculations  correspond to points 2 (recommended values), 5, 6,
8, and 9 in Fig.~2.

\begin{figure}[h]
\begin{center}
\includegraphics[height=3.5in]{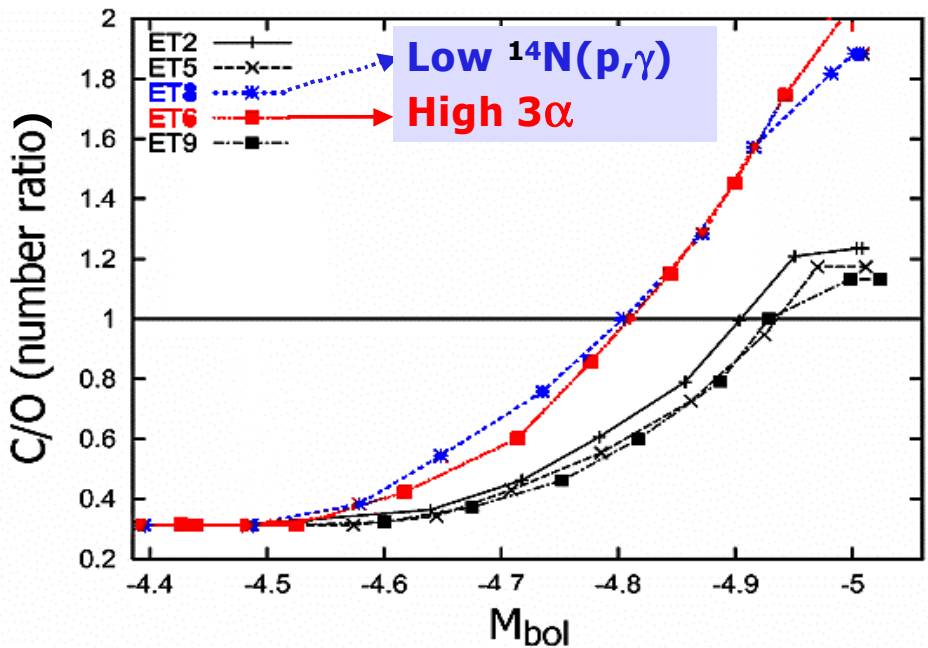}
\end{center}
\vspace{-0.4in} \caption{Values of the C/O ratio at the stellar
surface. Time increases to the right.  The values reached at the
end of the AGB phase are at the far right.}
\end{figure}

 The carbon enrichment of the stellar surface and the
production of $^{12}$C increased by about a factor of two when
either the $3\alpha$ rate was increased by its uncertainty (case
6), or when the $^{14}N(p, \gamma)^{15}$O rate was decreased by
its uncertainty (case 8).  The opposite changes (cases 5 and 9)
produced much smaller effects. These results are shown in more
detail in Fig.~3. The values for the two highest lying curves are
characteristic of  carbon stars. As noted previously they
correspond to cases 6 and 8 in Fig.~2.

Of course these results by themselves do not constitute a
resolution of the carbon star problem. Other poorly understood
phenomena might produce opposite changes \cite{herwig04}, or the
true rates might not be at the limiting values that lead to large
effects. But they do demonstrate that the nuclear reaction rates
must be better understood to reach a reliable theoretical
conclusion.

Fortunately, improved values of the  S factor $S_{1,14}$ for
$^{14}$N($p, \gamma)^{15}$O reaction have been obtained from new
experimental results from Duke, LUNA, Texas A and M, and Tokyo;
the resulting reaction rate is about half that recommended by
NACRE. Work of the LUNA group was presented at this Symposium
\cite{costantiniluna04}. A detailed summary of the various
experiments is given in \cite{herwig04}. Based on our evaluation
of these results we have chosen to use an unweighted average:
$S_{1,14}= 1.7\pm 0.25$ keV b. This value is smaller than the
lower limit of the NACRE values, and makes it more probable that a
large $^{12}$C production will result from the AGB process.

We have undertaken an extended set of calculations to investigate
this in more detail. Preliminary results have been obtained for
the left-hand points in Fig.~2 where we have  considered also
simultaneous variations of the two rates.   The calculation with
the smaller $^{14}$N($p, \gamma)^{15}$O rate indicated by the new
data (case 13) yields still larger $^{12}$C production. And a
simultaneous low $^{14}$N($p, \gamma)^{15}$O and high $3\alpha$
rate (cases 14 and 15) yields a further $^{12}$C enhancement. We
have not yet combined low  $^{14}$N($p, \gamma)^{15}$O  AND low
$3\alpha$ rates to find the result of this combination. It is
clear, however, that the AGB physics provides a further incentive
for obtaining accurate values of the $3\alpha$ rate.

\section{IMPROVING THE $\bf 3\alpha$ RATE}

For reference, the nuclei involved in the $3\alpha $ process are
shown in Fig 4.  As can be seen there, the process will be
strongest if the two resonances involved, the ground state of
$^8$Be and 7.65 MeV state of $^{12}$C, lie in the Gamow window.
For temperatures in the range encountered in the present
scenarios this is so, the process is resonant, and the rate of
the $3\alpha$ reaction has a simple form:
\begin{equation}\label{3alpha}
 r_{3\alpha}\propto \Gamma_{rad}\exp(-Q/kT).
\end{equation}

\begin{figure}[h]
\begin{center}
\includegraphics[height=4in]{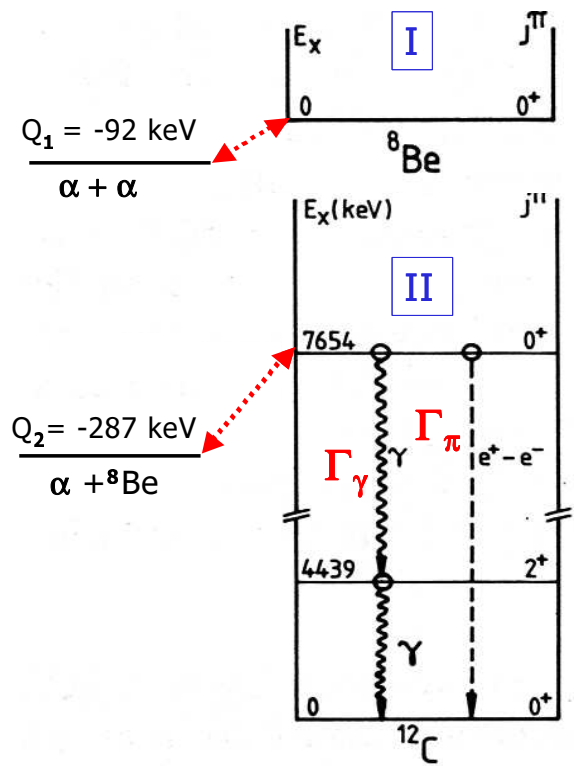}
\end{center}
\vspace{-0.4in}
 \caption{Nuclei involved in the $3\alpha $
process. The first step (I) of this process is the fusion of two
$\alpha $ particles to form a equilibrium concentration of $^8$Be.
The subsequent (II) capture of an alpha particle forms  an
equilibrium concentration of $^{12}$C in its 7.65 MeV state.
Occasionally $^{12}$C is formed by a leak, via a $\gamma $ cascade
or pair emission, to the ground state of $^{12}$C.}
\end{figure}
In this circumstance, only two quantities need to be determined:
$Q$ and $\Gamma_{rad}$.   $Q=E_r-3M_{\alpha}c^2$, where $E_{r}
\sim 7.6$ MeV and is known to $\pm 0.2$ keV from  measurements of
$E_{r}$~\cite{nolen76}. We note that the error quoted in many tabulations
of $^{12}$C levels is incorrect; it was taken from the wrong table
in \cite{nolen76}. The uncertainty due to this factor is
negligible, 1.2\% for $T_8=2$ and decreasing at higher T.
$\Gamma_{rad}$ is the radiative width of the 7.65 MeV state. At
higher and lower $T$ the situation is more complex \cite{angulo99}.
New experiments
\cite{fynbo03,fynbo04} provide no evidence for a 9.2 MeV $2^+$
state that is predicted theoretically \cite{angulo99} to have a
significant effect on $r_{3\alpha}$ at $T_9 > 2$.  Instead there
is evidence for a $0^+$ state at 11.2 MeV that will interfere with
the 7.6~MeV $0^+$ state and modify $ r_{3\alpha}$ at low
temperatures \cite{diget04}; details of these effects are not yet
published.

Essentially all the uncertainty in $r_{3\alpha}$ at the
temperatures relevant to the present considerations is due to the
uncertainty in $\Gamma_{rad}$ and is given by the product of three
quantities as shown in Fig.~5. All these quantities have been
measured several times with generally consistent results. This is
a tribute to the experimenters involved, since all these
measurements are difficult. Nevertheless, two quantities, the pair
width and the pair-decay branch of the 7.6~MeV state, need to be
measured better.

The pair width $\Gamma_{\pi}$ is determined from the transition
charge density for inelastic electron scattering to the 7.6~MeV
state. There is a new, as yet unpublished result \cite{crannell04},
based on a compendium of extant measurements over a large momentum
transfer range, that has a quoted accuracy of $\pm 2.7\%$. It is
difficult to imagine that a more accurate value of $\Gamma_{\pi}$
can be obtained. On the other hand, this value is not quite
consistent with the earlier values of $\Gamma_{\pi}$.

The pair branch $\Gamma_{\pi}/\Gamma$ is the least well known
quantity, primarily because it is so small, about $6\times 10^{-6}$.
A new experiment \cite{wuosmaa04}, a Western Michigan University
(WMU), Michigan State University (MSU) collaboration, is underway
using the Tandem accelerator at WMU.  A schematic of the apparatus
is shown in Fig.~6. This detector is an improved version of that
used by Robertson {\it et al.} \cite{robertson77}.
\begin{figure}[h]
\begin{center}
\includegraphics[width=6in]{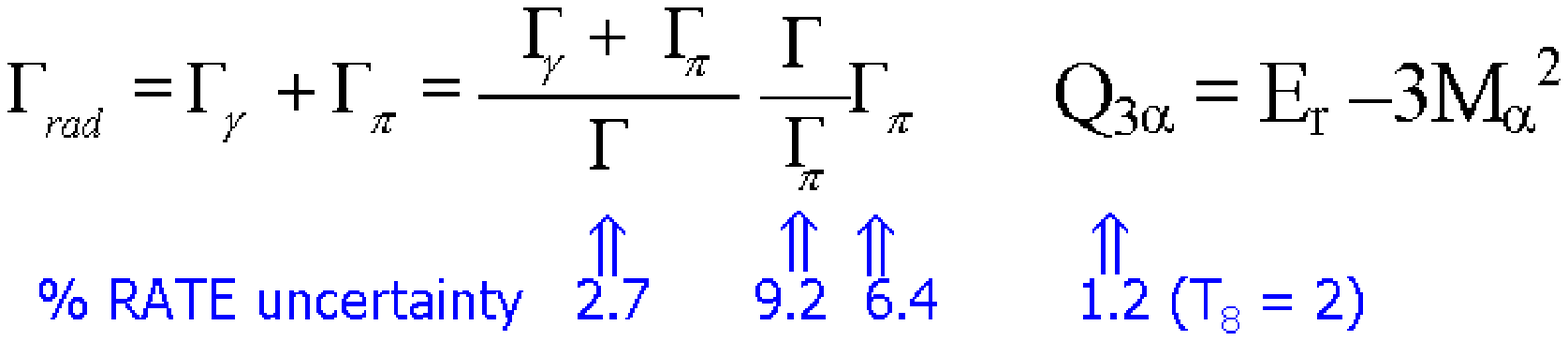}
\end{center}
\vspace{-0.4in} \caption{Expressions for  $\Gamma_{rad}$ and $Q$.
The uncertainties in $r_{3\alpha}$ from the various terms are also
shown.}
\end{figure}
The 7.6~MeV state in $^{12}$C  is excited by inelastic proton
scattering, taking advantage of a strong resonance at an
excitation energy of 10.6 MeV and a scattering angle of 135
degrees in the lab.  In order to reduce gamma ray backgrounds, a
coincidence is required between a thin cylinder and a large
plastic scintillator surrounding it; this should little affect the
number of detected pairs, but will strongly discriminate against
$\gamma $ rays--they have only small probability of interacting in
the thin cylinder.  The pair branch is given simply by the ratio
of the number of positron-electron pairs detected in the plastic
scintillators to the number of counts in the 7.65~MeV peak in the
proton spectrum. An examination of the systematic uncertainties in
the similar Robertson experiment leads us to estimate that an
accuracy of 5\% is achievable.

\begin{figure}[thb]
\begin{center}
\includegraphics[height=3.5in]{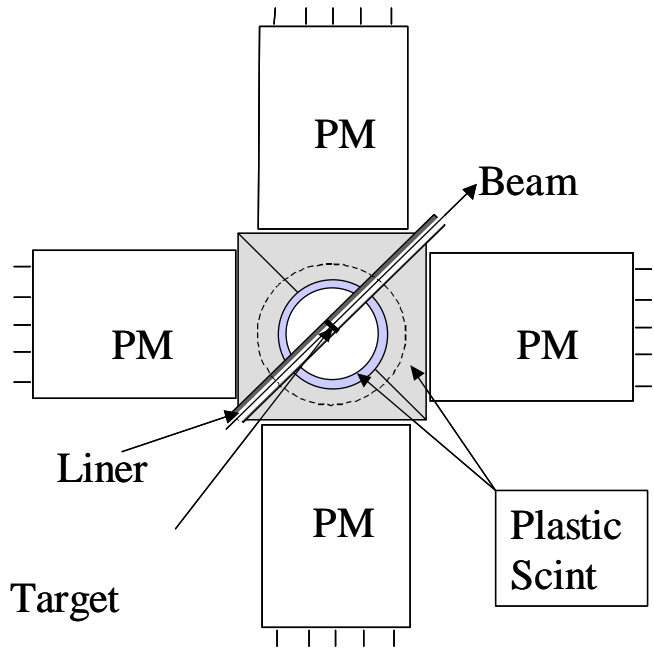}
\end{center}
\vspace{-0.4in} \caption{Schematic diagram of an apparatus for
observing the pair branch of the 7.65~MeV state in $^{12}$C.}
\end{figure}

\section{SUMMARY AND COMMENTS}
An examination of the effects of the rate for the $3\alpha$
process on the production of  $^{12}$C in core-collapse supernovae
(Type II) and in low mass AGB stars shows that this rate is
inadequately known. Variations within the $\pm 12\%$ errors cause
a change of $0.2M_{sun}$ in the core mass of a $25M_{sun}$
pre-supernova star and a factor of two change in the surface
abundance of $^{12}$C in light AGB stars.  In the supernova case,
it is the ratio of $r_{3\alpha}$ and $r_{1,14}$ that is important
at the 10\% level.  To reach this level, both these rates must be
significantly better known.  In the AGB case, only the $3\alpha$
rate is important, and significantly better accuracy is required.

It is likely that the uncertainty in the $3\alpha$ rate can be
halved by ongoing experiments and analyses.  Further improvements
beyond this would be extremely difficult.

A final comment concerns the nature of the core mass in
pre-supernova stars.  It appears that rather strong mass
enhancements can occur over a small range of R.  Thus it is
possible that the core collapse simulations use a mass that, for a
25 $M_{sun}$ star, is too large (or too small) by perhaps
$0.2M_{sun}$.  Such a difference might be large enough to cause an
explosion to fail or succeed just because the wrong model and core
mass were used.  It seems germane to ask whether sufficient
attention is being paid to the detailed nature of the
pre-supernova stars employed in supernova calculations.

\bibliography{nic8}

\begin{thebibliography}{10}

\bibitem{herwig04}
F. Herwig and Sam M. Austin, Astrophys. J. {\bf 613}, L73 (2004).

\bibitem{woosley03}
S. E. Woosley, A. Heger, T. Rauscher, and R.D. Hoffman, Nucl. Phys. {\bf A718},
  3c (2003).

\bibitem{buchmann96}
L. Buchmann, Astrophys. J. Lett. {\bf 468},127 (1996).

\bibitem{angulo99}
C. Angulo, {\it et al.}, Nucl. Phys. {\bf A690}, 765(1999).

\bibitem{costantiniluna04}
H. Costantini, for LUNA Collaboration, Abstract F04.

\bibitem{nolen76}
J. A. Nolen, Jr., and S. M. Austin, Phys. Rev. C {\bf 13}, 1773 (1976).

\bibitem{fynbo03}
H. O. U. Fynbo, {\it et al.}, Phys. Rev. Lett. {\bf 91}, 082502 (2003).

\bibitem{fynbo04}
H. O. U. Fynbo, Private communication, 2004.

\bibitem{diget04}
C. A. Diget, unpublished.

\bibitem{crannell04}
H. Crannell, {\it et al.}, Abstract F027, This Symposium.

\bibitem{wuosmaa04}
A. Wuosmaa, K. Starosta, and Sam M. Austin, unpublished.

\bibitem{robertson77}
R. G. H. Robertson, R. A. Warner, and S. M. Austin, Phys. Rev. C {\bf 15}, 1072
  (1977) .

\end{thebibliography}
\end{document}